\documentclass[a4paper]{jpconf}
\usepackage{graphicx}

\usepackage{hyperref}

\def\asec{\ifmmode ^{\prime\prime}\else$^{\prime\prime}$\fi}

\def\msun{M$_{\odot}$}

\def\it{\sl}
\def\degs{\ifmmode ^{\circ}\else$^{\circ}$\fi}
\def\amin{\ifmmode ^{\prime}\else$^{\prime}$\fi}
\def\asec{\ifmmode ^{\prime\prime}\else$^{\prime\prime}$\fi}
\def\fss{\hbox{$.\!\!^{\rm s}$}}        % Fractions of seconds
\def\farcs{\hbox{$.\!\!^{\prime\prime}$}}  % Fractions of arcseconds
\def\h{$^{\rm h}$}
\def\m{$^{\rm m}$}

\def\degs{\ifmmode ^{\circ}\else$^{\circ}$\fi}
\def\amin{\ifmmode ^{\prime}\else$^{\prime}$\fi}

\def\eqalign#1{\null\,\vcenter{\openup1\jot \m@th
   \ialign{\strut\hfil$\displaystyle{##}$&$\displaystyle{{}##}$\hfil
   \crcr#1\crcr}}\,}
\def\j1741{J1741+1351}
\def\psr{J1741}
\def\fer{\textit{Fermi}}
\def\msun{M$_\odot$}
\def\src{Source C}

\usepackage{xcolor}

\begin{document}
\title{Detection of the PSR J1741+1351 white dwarf companion with the Gran Telescopio Canarias}
\author{D A Zyuzin$^1$, 
A Yu Kirichenko$^2$,
A V Karpova$^1$, 
Yu A Shibanov$^1$,\\
S V Zharikov$^2$, 
E Fonseca$^3$ and 
A Cabrera-Lavers$^{4,5}$}

\address{$^1$Ioffe Institute, Politekhnicheskaya 26, St. Petersburg, 194021, Russia \\
$^2$Instituto de Astronom\'ia, Universidad Nacional Aut\'onoma de M\'exico, Apdo. Postal 877, Baja California, M\'exico, 22800  \\
$^3$Department of Physics \& McGill Space Institute, McGill University, 3600 University Street, Montreal, QC, H3A 2T8, Canada \\
$^4$Instituto de Astrof\'isica de Canarias, V\'ia L\'actea s/n, E38200, La Laguna, Tenerife, Spain\\
$^5$GRANTECAN, Cuesta de San Jos\'e s/n, E-38712, Bre\~{n}a Baja, La Palma, Spain
}

\ead{da.zyuzin@gmail.com}

\begin{abstract}

We report detection of the binary companion to the millisecond pulsar \j1741\
with the Gran Telescopio Canarias. 
The optical source position coincides with the pulsar coordinates
and its magnitudes are  $g'=24.84(5)$, $r'=24.38(4)$ and $i'=24.17(4)$.
Comparison of the data with the WD evolutionary models
shows that the source can be a He-core WD with a temperature of
$\approx 6000$ K and a mass of $\approx 0.2$\msun. 
The latter is in excellent agreement with the companion mass
obtained from the radio timing solution for PSR \j1741.
\end{abstract}

%%%%%%%%%%%%%%%%%%%%%%%%%%%%%%%%%%%%%%%%%%%%%%%%%%%%%%%%%%%
%%%%%%%%%%%%%%%%%%%%%%%%%%%%%%%%%%%%%%%%%%%%%%%%%%%%%%%%%%%
%%%%%%%%%%%%%%%%%%%%%%%%%%%%%%%%%%%%%%%%%%%%%%%%%%%%%%%%%%%
%%%%%%%%%%%%%%%%%%%%%%%%%%%%%%%%%%%%%%%%%%%%%%%%%%%%%%%%%%%

\section{Introduction}

Millisecond pulsars (MSPs) are neutron stars (NSs) that rotate particularly fast, having periods $P<30$ ms.
The canonical `recycling' scenario assumes that MSPs are formed in binary systems
where they spin up by transfer of mass and angular momentum from the secondary star
\cite{Bisnovatyi-Kogan1974, alpar1982}.

Optical observations of MSPs allow one to determine the nature and properties of their companions
\cite{vankerkwijk2005}.
This provides additional constraints on binary systems' parameters
which is important for understanding their formation and evolution.
In most cases, companions are low-mass white dwarfs (WDs) \cite{tauris2011}.
These objects are faint and thus hardly detectable.
Fortunately, the number of identifications gradually increases 
thanks to world's largest sky surveys and ground-based telescopes 
\cite{bassa2016,kirichenko2018,karpova2018}.

The binary MSP \j1741\ (hereafter \psr) was discovered with the Parkes radio telescope \cite{freire2006}
and then detected in $\gamma$-rays by \fer\ \cite{espinoza2013}.
It is among the best timed MSPs.
High-precision timing using 11-yr data set from the North American Nanohertz Observatory 
for Gravitational Waves provided measurements of the system inclination  
and masses of both the pulsar and its companion \cite{arzoumanian2018}.
The system parameters are presented in table~\ref{t:msp-par}.

To study the properties of the \psr\ companion, we performed
deep optical observations with the Gran Telescopio Canarias (GTC).
Here we report the results of the analysis of these data.

%%%%%%%%%%%%%%%%%%%%%%%%%%%% Timing solution %%%%%%%%%%%%%%%%%%%%%%%%%%%%
\begin{table}

\caption{Parameters of the PSR \j1741\ system obtained from \cite{arzoumanian2018}. 
The distance $D_p$ is derived from the timing parallax of 0.6(1) mas.
The dispersion measure distances $D_{\rm NE2001}$ and $D_{\rm YMW}$ were estimated
using the YMW16 \cite{ymw} and NE2001 \cite{ne2001} models, respectively. 
Numbers in parentheses are 1$\sigma$ uncertainties related to the last significant digits quoted.
}
\label{t:msp-par}
\begin{center}
\begin{tabular}{lc}
\br
Right ascension $\alpha$ (J2000)                                         & 17\h41\m31\fss144731(2) \\
Declination $\delta$ (J2000)                                             & +13\degs51\amin44\farcs12188(4) \\
\mr
Epoch (MJD)                                                              & 56209\\
Proper motion $\mu_\alpha =\dot{\alpha}{\rm cos}\delta$ (mas yr$^{-1}$)  & $-$8.98(2) \\
Proper motion $\mu_\delta$ (mas yr$^{-1}$)                               & $-$7.42(2) \\
\mr
Spin period $P$ (ms)                                                     & 3.747154500259940030(7) \\
Period derivative $\dot{P}$ ($10^{-20}$ s s$^{-1}$)                      & 3.021648(14) \\
Characteristic age $\tau$ (Gyr)                                          & 2\\
Orbital period $P_b$ (days)                                              & 16.335347828(4)\\
\mr
Dispersion measure (DM, pc cm$^{-3}$)                                    & 24.2 \\
Distance $D_{\rm NE2001}$ (kpc)                                          & 0.9 \\
Distance $D_{\rm YMW}$ (kpc)                                             & 1.36 \\
Distance $D_{p}$ (kpc)                                                   & $1.8^{+0.5}_{-0.3}$ \\\mr
Pulsar mass $M_p$ (\msun)                                                & $1.14^{+0.43}_{-0.25}$ \\
Companion mass $M_c$ (\msun)                                             & $0.22^{+0.05}_{-0.04}$ \\
System inclination $i$ (deg)                                             & $73^{+3}_{-4}$ \\
\br
\end{tabular}
\end{center}

\end{table}

%%%%%%%%%%%%%%%%%%%%%%%%%%%% Timing solution %%%%%%%%%%%%%%%%%%%%%%%%%%%%

%%%%%%%%%%%%%%%%%%%%%%%%%%%%%%%%%%%%%%%%%%%%%%%%%%%%%%%%%%%
%%%%%%%%%%%%%%%%%%%%%%%%%%%%%%%%%%%%%%%%%%%%%%%%%%%%%%%%%%%
%%%%%%%%%%%%%%%%%%%%%%%%%%%%%%%%%%%%%%%%%%%%%%%%%%%%%%%%%%%
%%%%%%%%%%%%%%%%%%%%%%%%%%%%%%%%%%%%%%%%%%%%%%%%%%%%%%%%%%%

\section{Observations and data reduction}
\label{sec:data}

The observations of the \psr\ field\footnote{Proposal GTC4-18AMEX, PI A. Kirichenko} 
were carried out in June 2018 in the Sloan $g'$, $r'$ and $i'$ bands with 
the GTC/OSIRIS\footnote{\url{http://www.gtc.iac.es/instruments/osiris}} instrument.
Dithered science frames 
were taken, with total exposure times of 
$3$, $3.5$  and $2.76$~ks for the $g'$, $r'$ and $i'$ filters, respectively. 
A short 20~s exposure was obtained in the $r'$ band to avoid saturation 
of bright stars that were further used for precise astrometry. The formal
$rms$ uncertainties of the astrometric fit are 0.05 arcsec in RA and Dec for all bands. 
The data reduction was performed with the Image Reduction and Analysis Facility (IRAF) package.
For the photometric calibration, we used the 
standard stars SA 112\_805 and SA 104\_428 
observed during the same night as the target. 
The derived photometric zeropoints are 
$z_{g'}=28.76(2)$, $z_{r'}=28.99(1)$ and $z_{i'}=28.55(1)$.
The values were obtained by comparing the standard star magnitudes from \cite{smith2002} with 
their instrumental magnitudes, corrected for the finite aperture and atmospheric extinction. 
We used the atmospheric extinction coefficients\footnote{\url{http://www.iac.es/adjuntos/cups/CUps2014-3.pdf}} 
$k_{g'}=0.15(2)$, $k_{r'}=0.07(1)$ and  $k_{i'}=0.04(1)$.

%%%%%%%%%%%%%%%%%%%%%%%%%%%%%%%%%%%%%%%%%%%%%%%%%%%%%%%%%%%
%%%%%%%%%%%%%%%%%%%%%%%%%%%%%%%%%%%%%%%%%%%%%%%%%%%%%%%%%%%
%%%%%%%%%%%%%%%%%%%%%%%%%%%%%%%%%%%%%%%%%%%%%%%%%%%%%%%%%%%
%%%%%%%%%%%%%%%%%%%%%%%%%%%%%%%%%%%%%%%%%%%%%%%%%%%%%%%%%%%

%%%%%%%%%%%%%%%%%%%%%%%%%%%%% Pulsar field %%%%%%%%%%%%%%%%%%%%%%%%%%%%%%%%%%
\begin{figure}

\begin{minipage}[h]{0.495\linewidth}
\center{\includegraphics[width=1.0\linewidth,trim={0 0 0 2cm},clip]{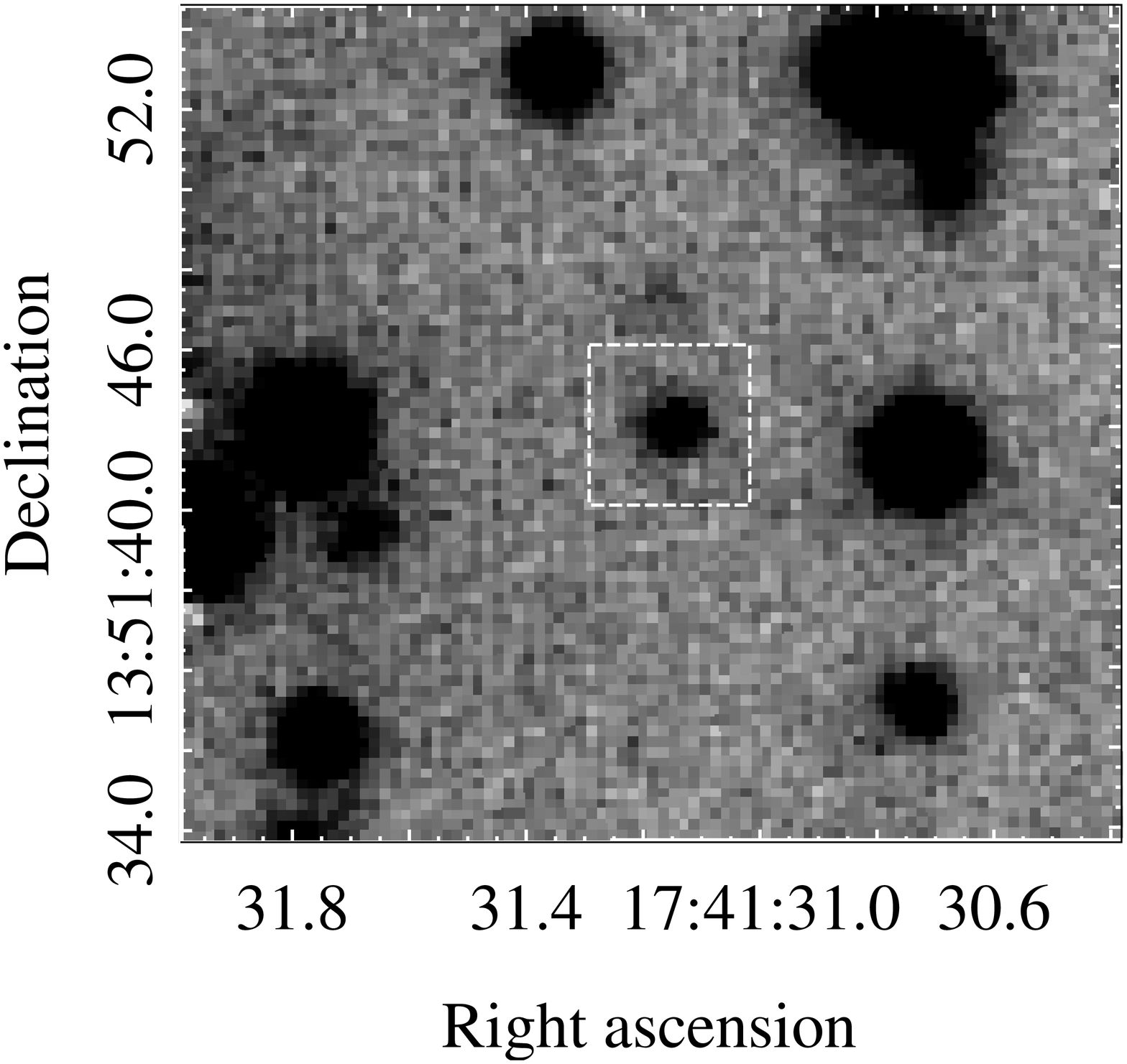}}
\end{minipage}
\begin{minipage}[h]{0.495\linewidth}
\center{\includegraphics[width=0.6\linewidth,clip]{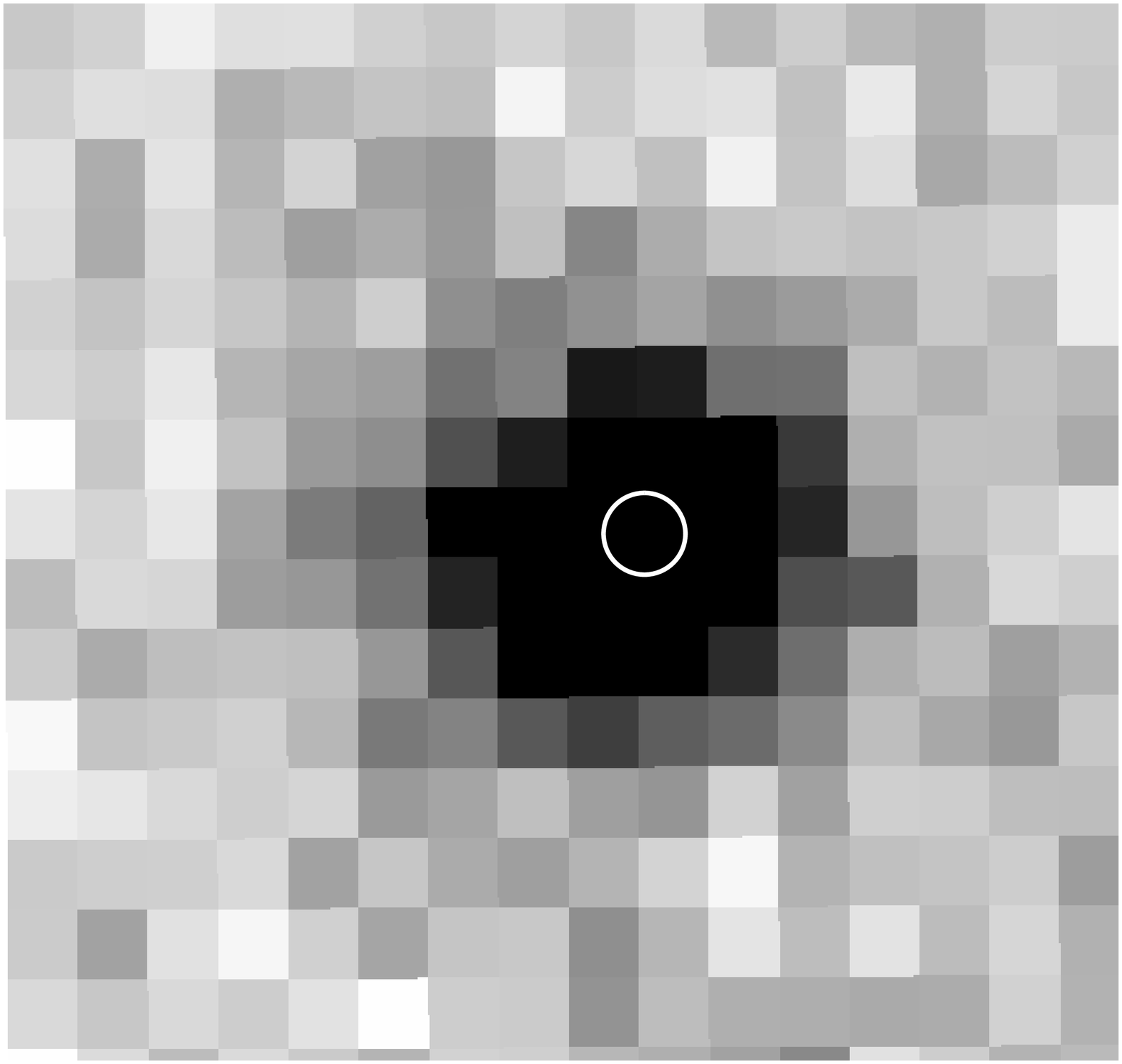}}
\end{minipage}
\caption{20\asec$\times$20\asec\ GTC/OSIRIS $r'$-band image of the \psr\ field (left) and 
its enlarged 4\asec$\times$4\asec\ section within the dashed rectangle containing the pulsar (right). 
The solid 
circle with a radius of 0\farcs15 shows the 3$\sigma$ position uncertainties of the pulsar radio position.
}
\label{fig:images}
\end{figure}
\vspace{-0.25cm}
%%%%%%%%%%%%%%%%%%%%%%%%%%%%% Pulsar field %%%%%%%%%%%%%%%%%%%%%%%%%%%%%%%%%%

\section{Results and discussion}
\label{sec:results}

%%%%%%%%%%%%%%%%%%%%%%%%%%%%%   Diagrams   %%%%%%%%%%%%%%%%%%%%%%%%%%%%%%%%%%
\begin{figure}
\begin{minipage}[h]{0.485\linewidth}
\center{\includegraphics[width=1.0\linewidth, clip=]{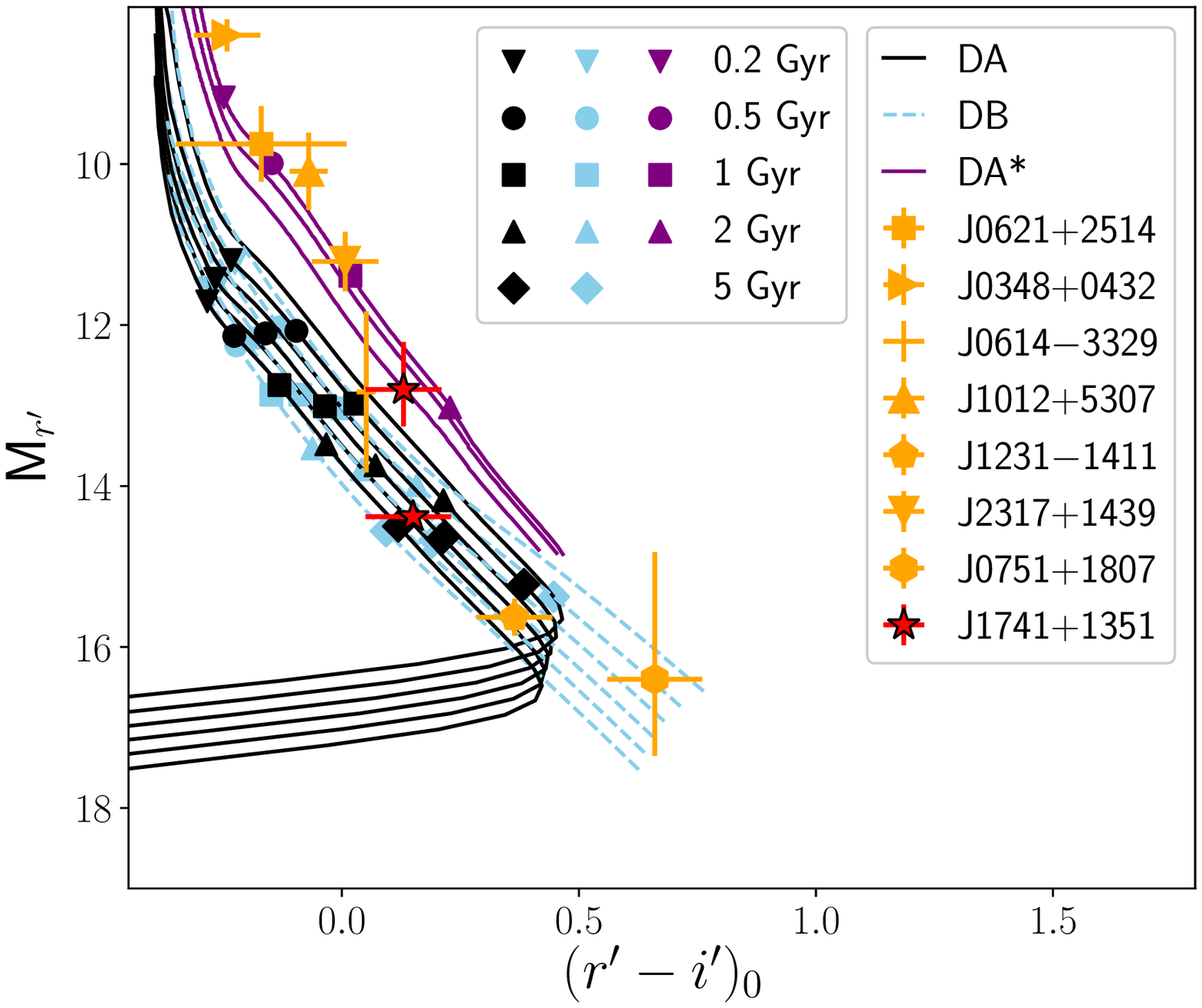}}
\end{minipage}
\begin{minipage}[h]{0.485\linewidth}
\center{\includegraphics[width=0.92\linewidth, clip=]{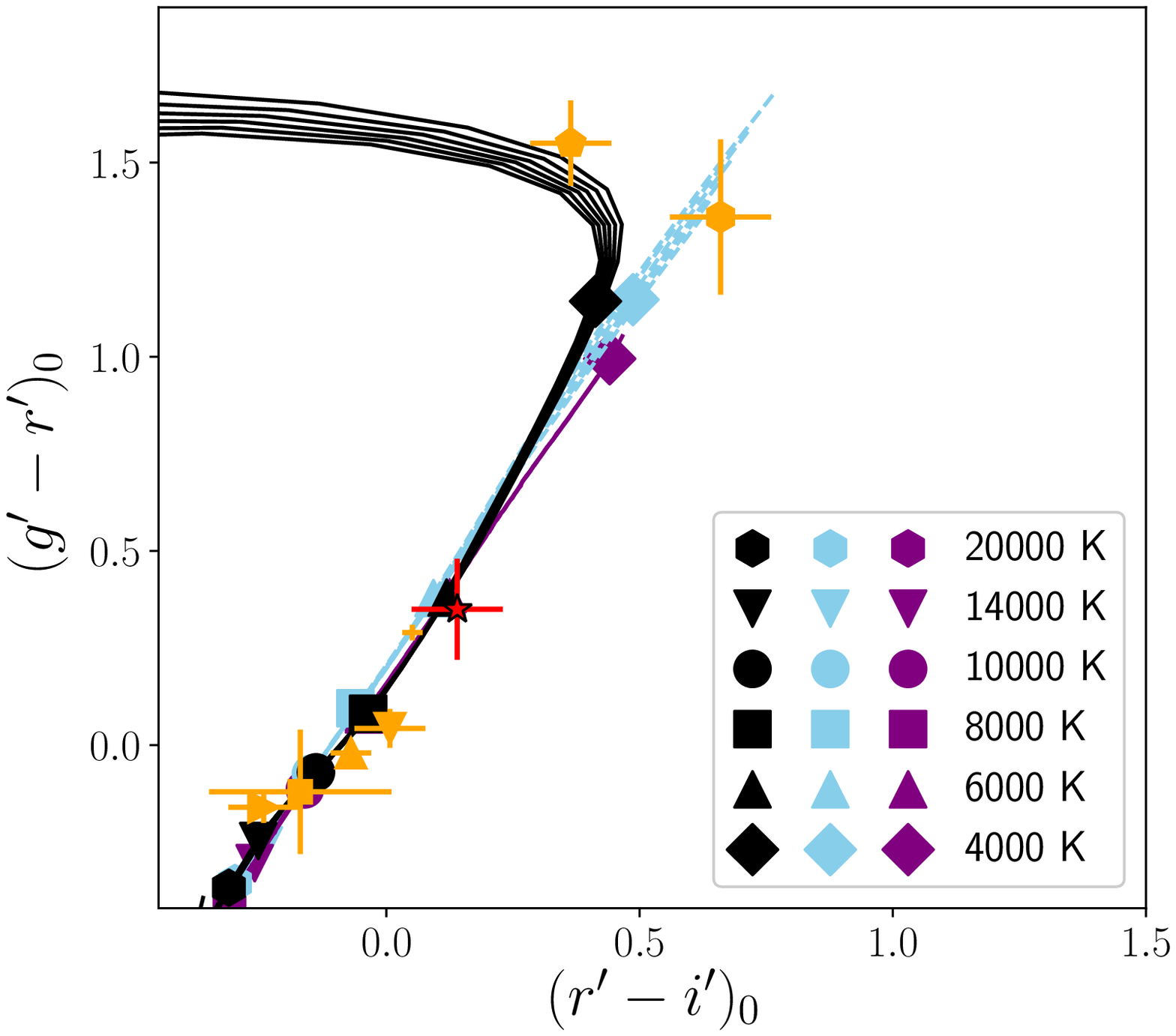}}
\end{minipage}
\caption{Colour-magnitude (left) and colour-colour (right) diagrams with various WD cooling tracks.  
Solid black (DA) and dashed light-blue (DB) lines show the cooling tracks for WDs 
with hydrogen and helium atmospheres, respectively 
\cite{holberg2006,kowalski2006,tremblay2011,bergeron2011}, 
with masses 0.3--0.8~${\rm M_\odot}$ (spaced by $0.1~{\rm M_\odot}$) increasing from upper to lower curves.
Solid purple (DA*) lines show the models from \cite{panei2007} for WDs with hydrogen
atmospheres and masses of 0.1869, 0.2026 and 0.2495 \msun.
Cooling ages in the left panel and temperatures in the right panel 
are indicated by different symbols.
The location of the \psr\ presumed companion is marked by the red star: the upper one is for the maximum distance estimate $D_p$ and the lower one is for
the minimum distance estimate $D_{\rm NE2001}$ kpc. 
The data for WD companions of other pulsars listed in the legend are also shown
(see \cite{beronya2019} for details).
}
\label{fig:col-mag}
\end{figure}
%%%%%%%%%%%%%%%%%%%%%%%%%%%%%   Diagrams   %%%%%%%%%%%%%%%%%%%%%%%%%%%%%%%%%%

In figure~\ref{fig:images} we show sections of the \psr\ field
in the $g'$, $r'$ and $i′$ bands. 
The pulsar coordinates at the epoch of our observations (MJD 58274, $i'$ filter) 
corrected for its proper motion (table~\ref{t:msp-par}) are
$\alpha$ = 17\h41\m31\fss141245(8) and $\delta$ = +13\degs51\amin44\farcs0799(1).
Its position uncertainty is shown in the images by the circle which also
accounts for the astrometric referencing uncertainty. 
In all bands, we detect a starlike source (hereafter \src) 
overlapping with the pulsar position.

We obtained the \src~magnitudes $g'=24.84(5)$, $r' = 24.38(4)$ and $i'=24.17(4)$.
To correct these values for the interstellar extinction,
we used the dust map \cite{green2018} 
and distances to \psr\ (table~\ref{t:msp-par}).
We got the reddening $E(B-V)$ of 
$0.10(3)$ 
and 
$0.13(2)$ for the minimum and maximum distance estimates.
$E(B-V)$ was then transformed to the extinction correction values
using coefficients from \cite{schlafly2011}.
For $D_{\rm NE2001}=0.9$ kpc, the resulting intrinsic colours and absolute magnitude
are 
$(g'-r')_0 = 0.36(14)$, $(r'-i')_0 = 0.15(10)$ and 
$M_{r'} = 14.38(8)$ while for $D_p=1.8^{+0.5}_{-0.3}$ kpc -- 
$(g'-r')_0 = 0.33(11)$, $(r'-i')_0 = 0.13(8)$ and
$M_{r'} = 12.80^{+0.46}_{-0.59}$.

To check whether \src\ is a WD, we compared its colours and magnitudes
with the WD cooling tracks from 
\cite{panei2007,holberg2006,kowalski2006,tremblay2011,bergeron2011}\footnote{\url{http://www.astro.umontreal.ca/~bergeron/CoolingModels/}} 
which are shown in figure~\ref{fig:col-mag}.
Indeed, according to the diagrams, \src\ is likely a WD  
with a temperature of about 6000 K.
This estimate is appropriate for the whole range of distance estimates
to \psr\ (table~\ref{t:msp-par}) since the reddening 
does not vary significantly. 
However, various distances imply different estimates of other parameters of \src.
The minimum distance corresponds to a WD with a mass of $>0.4$ \msun\
and age of 2--5 Gyr and the maximum one -- to a WD with a mass of about 0.2--0.3 \msun\ and age of 1--2 Gyr.
The former case is not compatible with radio timing measurements of 
the \psr\ companion mass 
$M_c=0.22(5)$ \msun\
while the latter one is in a good agreement with it.
This supports the larger estimate of the distance to \psr~and implies that its companion is a DA He-core WD.

\ack{The manuscript is based on observations made with the GTC, installed in the Spanish Observatorio
del Roque de los Muchachos of the Instituto de Astrof\'isica de Canarias, in the island of La Palma.
We thank the NANOGrav collaboration for sharing data products from the 12.5-yr data set.
DAZ thanks Pirinem School of Theoretical Physics for hospitality.
AYK and SVZ acknowledge support from DGAPA/PAPIIT Project IN100617.}

%%%%%%%%%%%%%%%%%%%%%%%%%%%%%%%%%%%%%%%%%%%%%%%%%%%%%%%%%%%
%%%%%%%%%%%%%%%%%%%%%%%%%%%%%%%%%%%%%%%%%%%%%%%%%%%%%%%%%%%
%%%%%%%%%%%%%%%%%%%%%%%%%%%%%%%%%%%%%%%%%%%%%%%%%%%%%%%%%%%
%%%%%%%%%%%%%%%%%%%%%%%%%%%%%%%%%%%%%%%%%%%%%%%%%%%%%%%%%%%

\section*{References}

\bibliography{refmsp}

\end{document}